\documentclass[11pt]{article}
\usepackage{amsmath,amsfonts,amssymb,amsthm}
\usepackage[margin=1in]{geometry}
\usepackage[colorlinks]{hyperref}
\usepackage{graphicx}

\begin{document}


\title{Non-maximally entangled mixed states of $X$ and non-$X$ types as teleportation channels}

\author{Anushree Bhattacharjee$^{1}$, Sovik Roy$^{2}$
\thanks{a.bhattacharya.tmsl@ticollege.org} 
\thanks{s.roy2.tmsl@ticollege.org}\\
$^1$ $^2$  Department of Mathematics, Techno Main Salt Lake (Engg. Colg.), \\Techno India Group, EM 4/1, Sector V, Salt Lake, Kolkata  700091, India\\\\
Md. Manirul Ali$^{3}$
\thanks{manirul@citchennai.net}\\
$^3$  Centre for Quantum Science and Technology, \\Chennai Institute of Technology, Chennai 600069, India\\\\
Biplab Ghosh$^{4}$ 
\thanks{drbiplabghosh@vivekanandacollegeforwomen.org}\\
$^4$ Department of Physics, \\Vivekananda College for Women, Barisha, Kolkata - 700008, India
}

\maketitle

\begin{abstract}
\noindent 
Mixed spin-$\frac{1}{2}$ states violating Bell-CHSH inequality is useful for teleportation. There exist states which do not violate Bell-inequality but is still useful as teleportation channels.  Maximally entangled mixed states of Munro class and Ishizaka-Hiroshima class are such types which although satisfy Bell-CHSH inequality, yet can perform better as teleportation channels for a given degree of mixedness\cite{adhikari2010}. In this work we construct class of mixed states of non-maximally entangled types whose efficacy as teleportation channels have been studied. For certain range of state parameters, these non-maximally entangled mixed states performs better as quantum teleportation channels than certain maximally entangled mixed states (such as Werner state). These constructed states, though entangled, satisfy Bell-CHSH inequality implying further that violation of local inequalities may not be good indicators of their ability to complete quantum processing tasks such as teleportation.

Pacs No.: 03.65.Ud ~~\and \and ~~03.67.-a
\end{abstract}


\section{Introduction:}\label{sec:introduction}
Recent progresses in the field of quantum information science has marked significant strides, advancing the comprehension in the quantum environment. The revolution in quantum information science is propelled by quantum tasks such as quantum teleportation \cite{bennett1993}, super dense coding \cite{bennett1992}, quantum key distribution \cite{bennett1984,ekert1991}, quantum computation \cite{deutsch1983,jozsa1998,shor1994} et. al. Among the above quantum phenomena, quantum teleportation has garnered widespread attention among scientific researchers. Being a non-classical technique for transmitting quantum information across significant distances, quantum teleportation stands out as a highly pertinent application in the realm of quantum information processing. The concept of teleportation was first introduced by Bennett et al \cite{bennett1993}. Dik Bouwmeester et. al. successfully executed the experimental implementation of quantum teleportation \cite{bouwmeester1997}. Later Popescu found that the pairs of mixed states can also be used for teleportation, a feature known as probabilistic teleportation or imperfect teleportation \cite{popescu1994}.  Teleportation requires the separation of a protocol into classical and quantum parts using which an unknown state is reconstructed with perfect fidelity at another location. In this process the original copy of the unknown qubit is destroyed. Not all quantum states are efficient as quantum teleportation channels. This restriction is captured by the measure known as teleportation fidelity \cite{popescu1994,massar1995,gisin1996}. Classically the maximum possible fidelity achievable by means of local operations and classical communication (LOCC) is $\frac{2}{3}$. For a successful achievement of quantum teleportation, the fidelity of the quantum channel must exceed this classical fidelity. Efficiency of teleportation reaches its peak only when the sender and the receiver share maximally entangled pure state. The assessment of teleportation efficiency is quantified by a parameter known as fidelity, which takes the value $1$ for maximally entangled states and $0$ for separable states. Almost all teleportation schemes utilize maximally entangled states which theoretically ensure the attainment of highest fidelity, however in practical situations, environmental interactions lead to a gradual degradation of entanglement between particles. Hence, in ideal scenario, pure maximally entangled states is anticipated to yield a high fidelity performance in achieving a task whereas, the preparation of pure states presents a formidable challenge. Instead, pragmatically, states often dealt with, exhibit mixed characteristics. Quantum states that attain greatest possible entanglement relative to a given degree of mixedness are classified as Maximally Entangled Mixed States (MEMS) \cite{ishizaka2000}, while states that do not reach the same pinnacle of entanglement  are categorized as Non-Maximally Entangled Mixed States (NMEMS) \cite{hiroshima2000}. Werner state characterized as a maximally entangled mixed state proved to be a valuable resource in the context of teleportation \cite{adhikari2010}. Munro et. al.  investigated a class of states characterized by the utmost entanglement achievable for a specific level of purity and formulated an analytical expression to describe that class of MEMS \cite{munro2001}, which can be termed as MJWK type states. Adhikari et. al. demostrated that the MEMS proposed by MJWK  can be used as a reliable teleportation channel when mixedness of the state  falls below the critical threshold value of $\frac{22}{27}$. However, Werner state outperforms MJWK MEMS class with respect to its capacity as quantum teleportation channel. They also studied the bounds of teleportation fidelity of Werner state for which it can be used as teleportation channel. Adhikari et. al.  also proposed a new class of bipartite non-maximally entangled mixed state by taking convex combination of GHZ state and W state. This state is more efficient as teleportation channel than Werner derivative class of states (which is a different type of NMEMS, obtained by applying a non-local unitary operator on the Werner state)  \cite{adhikari2010}\footnote{Two of the authors S. Roy and B. Ghosh of this work were also the co-authors of the paper.}.\\\\
In this paper we construct a few special bipartite mixtures and study their efficiency as quantum teleportation channels. These states are of NMEMS types and their efficiency as quantum teleportation channel have been ascertained. For this, we have first considered tripartite pure states such as $\vert W\rangle,\: \vert \bar{W}\rangle,\: \vert W\bar{W}\rangle$ and $\vert Star\rangle$ \cite{cao2020}. We then remove (or trace out) a party from these pure states and take convex combination of these bipartite mixtures (derived from tripartite states) with two qubit maximally entangled Bell state. $\vert W\bar{W}\rangle$ state, characterized by an equal superposition of a standard $\vert W\rangle$ state and its spin-flipped version $\vert \bar{W}\rangle$ state, is significant because it has quantum correlations at both bipartite and tripartite levels. Such a state is a good test bed for studying quantum correlations distributed at different levels. The presence of different types of correlations is one of the reasons that $\vert W\rangle$ states are robust under local decoherence. Similar to $\vert W\bar{W}\rangle$ state, the $\vert Star\rangle$ state also exhibits distribution of correlations at all conceivable levels, but $\vert Star\rangle$ state introduces an intriguing contrast to $\vert W\bar{W}\rangle$ state by showcasing correlations in an asymmetric manner. This is because, in $\vert Star\rangle$ state two types of qubits are present, viz, central and peripheral qubits. Central qubit is that qubit which when traced out, the remaining two qubits are left in separable state. Those two left out qubits, however, are peripheral qubits. Recently Roy et.al. investigated both $\vert W\bar{W}\rangle$ and $\vert Star\rangle$ states and explored different quantum properties with respect to both coherent and incoherent basis \cite{roy2023}. \\\\
The current paper is organised as follows. In sec.\ref{sec:measures} we have discussed briefly, the measures used in quantum information science such as to study how entangled the states are, quantifying mixedness of the state, how to verify whether states are useful as quantum teleportation channels and to know the signature status of entanglement of the constructed state by the violation of Bell-inequality. In sec.\ref{sec:bipartite} we build the bipartite mixtures which are non-maximally entangled mixed states. We find that, among our constructed states some are $X$ states and others are not of $X$ type, which, in this article, are termed as non-$X$ states. We, in sec.\ref{sec:measures1}, study the teleportation fidelity, Bell-CHSH inequality violation, nature of entanglement and mixedness of the $X$ type NMEMS while in sec.\ref{sec:nonX}, we study the same for those states that are of non-$X$ type. Sec.\ref{sec:conclusion} gives concluding remarks.
\section{Quantum measures : A brief overview}
\subsection*{Concurrence, teleportation fidelity and linear entropy:}{\label{sec:measures}}
Entanglement is a fundamental quantum resource emerging from the non-local correlations among quantum systems. Among the various measures such as concurrence, relative entropy of entanglement, negativity and log negativity et. al. that have been developed to quantify entanglement, concurrence stands out as universally recognized measure for two qubit system, with equal efficacy in assessing entanglement for both pure and mixed quantum states \cite{wk1998}. The concurrence for a bipartite state $\rho$ is defined as 
\begin{eqnarray}
\label{conc1}
C(\rho)= max \{0, \sqrt{\lambda_{1}}-\sqrt{\lambda_{2}}-\sqrt{\lambda_{3}}-\sqrt{\lambda_{4}}\}
\end{eqnarray}
where $\lambda_{i}, (i = 1, 2, 3, 4)$ are the eigenvalues of the matrix $\rho\tilde{\rho}$ in decreasing order in which $\tilde{\rho}$ is the spin-flipped density matrix defined as \\
\begin{equation}
\tilde{\rho}=(\sigma_{y}\otimes \sigma_{y}){\rho^{*}}(\sigma_{y}\otimes \sigma_{y})
\end{equation}
where ${\rho^{*}}$ is the complex conjugate of the density matrix $\rho$ and $\sigma_{y}$ is the Pauli spin operator.\\\\
For an arbitrary density matrix $\rho$, when used as quantum teleportation channel, it's optimal teleportation fidelity is given by
\begin{eqnarray}
\label{telep1}
f^{T}(\rho) = \frac{1}{2}\Big\lbrace 1+ \frac{N(\rho)}{3} \Big\rbrace,
\end{eqnarray}
where $N(\rho) = \sum_{i = 1}^{3}\sqrt{u_{i}}$. Here $u_{i}$'s are the eigenvalues of the matrix $T^{\dagger}T$. The elements of $T$ are denoted by $t_{nm}$ and these elements are calculated as $t_{nm} = Tr(\rho\sigma_{n}\otimes\sigma_{m})$ where $\sigma_{j}$'s denote the Pauli spin operators. In terms of teleportation fidelity, a general result holds that any mixed spin-$\frac{1}{2}$ state $\rho$ is useful for standard teleportation if and only if $N(\rho)>1$ \cite{horohorohoro1996}.
\\\\
When interacted with external environment, the quantum states get affected and there is degradation of quantumness of these states. This phenomenon is known as decoherence and such a state is called mixed state. Mixed states can be probabilistically used as quantum teleportation channel and their mixedness are also studied. Linear entropy is one such measure to quantify the mixedness of a given mixed state. For an arbitrary $d-$ dimensional quantum mixed state with density matrix $\rho$, the mixedness, using the normalized linear entropy, is defined as
\begin{eqnarray}
\label{le1}
L(\rho) = \frac{d}{d-1}(1-Tr(\rho^{2})),
\end{eqnarray}
where $Tr(\rho^{2})$ describes the purity of the quantum system \cite{peters2004}. For bipartite system, the value of $L$ ranges from $0$ to $1$. For any pure state $\rho$, $L(\rho) = 0$ and when $\rho = \frac{I}{4}$, i.e. when $\rho$ is maximally mixed state, then $L(\rho)$ is $1$.
\subsection*{Bell-CHSH inequality violation:}
Any quantum state described by the density operator $\rho$ violates the Bell-CHSH inequality if and only if the following condition is satisfied i.e.
\begin{eqnarray}
\label{bell-chsh}
M(\rho) = \max_{i>j}(u_{i}+v_{j})>1.
\end{eqnarray}
Here $u_{i}$'s are the eigenvalues of the matrix $T^{\dagger}T$.\\\\
Moreover, Horodecki et.al \cite{horohorohoro1996} had shown that every mixed two spin-$\frac{1}{2}$ state which violates any generalized Bell-CHSH inequality is useful for teleportation. If we denote by $P$ the statement that $``$ $\rho$ violates Bell-CHSH inequality" and by $Q$ we mean $``$ the state $\rho$ is useful for teleportation", then by Horodecki's paper we have $``$ $P$ $\Rightarrow$ $Q$". By mathematical logic one can immediately conclude that $``$ $\neg Q$ $\Rightarrow$ $\neg P$". This means that the state which is not useful for teleportation will satisfy Bell-CHSH inequality. But one cannot immediately conclude that $``$ $Q$ $\Rightarrow$ $P$", i.e. there may exist states which satisfy Bell-CHSH inequality, but still can be used as teleportation channel. In this context we are now going to construct some classes of non-maximally entangled mixed states to study how these states behave according to Bell violation and whether they are efficient as quantum teleportation channels.
\section{States: Bipartite Mixtures}
\label{sec:bipartite}
We know that the Bell states are a set of maximally entangled two qubit pure states and they are as follows: 

\begin{eqnarray}
\label{bell}
\vert \phi^{\pm}\rangle = \frac{1}{\sqrt{2}}\lbrace \vert 00\rangle \pm \vert 11\rangle\rbrace;   \qquad  \qquad \qquad
\vert \varphi^{\pm}\rangle = \frac{1}{\sqrt{2}}\lbrace \vert 01\rangle \pm \vert 10\rangle\rbrace.
\end{eqnarray}
Here $A$ and $B$ denote the two parties Alice and Bob holding one qubit each. On the other hand the tripartite $\vert W\rangle$ state and it's spin-flipped version $\vert\bar{W}\rangle$ are respectively defined as
\begin{eqnarray}
\label{w1}
\vert W\rangle &=& \frac{1}{\sqrt{3}}(\vert 001\rangle + \vert 010\rangle + \vert 100\rangle),\nonumber\\
\vert \bar{W}\rangle &=& \frac{1}{\sqrt{3}}(\vert 110\rangle + \vert 101\rangle + \vert 011\rangle),
\end{eqnarray}
where each party holds qubits individually. $\vert W\bar{W}\rangle$ state is characterized by an equal superposition of a standard $\vert W\rangle$ state and its spin-flipped version $\vert \bar{W}\rangle$ state. 
\begin{eqnarray}
\label{wwbar}
\lvert W\bar{W}\rangle=\frac{1}{\sqrt{2}}(\lvert W\rangle+\lvert\bar{W}\rangle)
\end{eqnarray}
Such a state serves as an excellent experimental platform for investigating quantum correlations distributed across varying levels. The existence of diverse correlations in $\vert W\rangle$ state maintains the robustness of the state under local decoherence. Another tripartite state that has been considered here is $\vert Star\rangle$, which is defined as
\begin{eqnarray}
\label{star}
\vert Star \rangle = \frac{1}{2}(\vert 000 \rangle + \vert 100 \rangle + \vert 101\rangle + \vert 111 \rangle).
\end{eqnarray}
Now from $\vert W\rangle$ (eq.(\ref{w1})) state tracing out any of the qubits and then taking convex combination of the reduced density matrix with that of four Bell states described in eq.(\ref{bell}), some new class of states are obtained. Similar construction procedures have been carried by taking into consideration the $\vert Star\rangle$ state (eq.(\ref{star})) and the Bell states. 
\subsection*{Class $1$ states:}
We introduce a distinct class of bipartite states by initially considering a $3-$ qubit state and subsequently removing the third party. The resulting two qubit density matrix $\rho^{W}$ is then combined through convex combination with each of four Bell states (eq.(\ref{bell})). The resultant class of states are then written as,
\begin{eqnarray}
\label{class1a}
\rho^{(1)}&=& p\rho^{W} + (1-p)\rho^{\phi^\pm},\nonumber\\
\rho^{(2)}&=& p^{\prime}\rho^{W} + (1-p^{\prime})\rho^{\varphi^\pm}, \: 0\leq p, p^{\prime} \leq 1.
\end{eqnarray}
Similarly by taking $\vert \bar{W}\rangle$ and $\vert W\bar{W}\rangle$ states, removing third party as before and taking convex combination with each of the four Bell states of eq.(\ref{bell}) we obtain the following class of states.
\begin{eqnarray}
\label{class1b}
\rho^{(3)}&=& q\rho^{\bar{W}} + (1-q)\rho^{\phi^\pm},\nonumber\\
\rho^{(4)}&=& q^{\prime}\rho^{\bar{W}} + (1-q^{\prime})\rho^{\varphi^\pm}, \: 0\leq q, q^{\prime} \leq 1.
\end{eqnarray}
and
from the symmetric structures of the $\vert W\rangle$ and $\vert \bar{W}\rangle$ states, we can also observe that removing qubits respectively held by first two parties will generate the same set of states as above.\\\\
In the limiting case where $p(or \:p^{\prime}) = 1$ and $q(or \:q^{\prime})=1$ in eqs.(\ref{class1a}) and (\ref{class1b}), the states $\rho^{(1)}(or \:\rho^{(2)})$ and $\rho^{(3)}(or \:\rho^{(4)})$ reduce to the states
\begin{eqnarray}
\label{mems1}
\rho^{W} &=& \frac{1}{3}\vert 00\rangle\langle 00\vert + \frac{2}{3}\vert \psi^{+}\rangle\langle \psi^{+}\vert,\nonumber\\
\rho^{\bar{W}} &=& \frac{1}{3}\vert 11\rangle\langle 11\vert + \frac{2}{3}\vert \psi^{+}\rangle\langle \psi^{+}\vert.
\end{eqnarray}
These states, as shown in (\ref{mems1}), are maximally entangled mixed states (MEMS) as they fall into the category of Ishizaka-Hiroshima's proposed class of MEMS \cite{ishizaka2000}. However, the states represented in eqs.(\ref{class1a}) and (\ref{class1b}) are non-maximally entangled mixed states (NMEMS).
\subsection*{Class $2$ states:}
Although the structures of $\vert W\rangle$ and $\vert \bar{W}\rangle$ are similar, $\vert W\bar{W}\rangle$ is different (which we will shortly emphasize) and removing third qubit, we construct a mixture by taking convex combination of $\rho_{AB}^{W\bar{W}}$ and the four possible Bell states (eq.(\ref{bell})) as follows.
\begin{eqnarray}
\label{class1c}
\rho^{(5)}&=& r\rho^{W\bar{W}} + (1-r)\rho^{\phi^\pm},\nonumber\\
\rho^{(6)}&=& r^{\prime}\rho^{W\bar{W}} + (1-r^{\prime})\rho^{\varphi^\pm}, \: 0\leq r, r^{\prime} \leq 1.
\end{eqnarray}
However, similar to the states $\vert W\rangle$ and $\vert \bar{W}\rangle$, the removal of first two qubits will yield similar states as defined in eq.(\ref{class1c}).\\\\
Considering the $\vert Star\rangle$ state defined in eq.(\ref{star}), we remove first qubit i.e. the peripheral qubit and take convex combination with each of the Bell states (eq.(\ref{bell})) to obtain the following.
\begin{eqnarray}
\label{class2a}
\tau^{(1)}&=& s\rho^{star} + (1-s)\rho^{\phi^\pm},   \nonumber\\
\tau^{(2)}&=& s^{\prime}\rho^{star} + (1-s^{\prime})\rho^{\varphi^\pm}, \: 0\leq s, s^{\prime} \leq 1.
\end{eqnarray}
In eqs.(\ref{class1a}), (\ref{class1b}), (\ref{class1c}) and (\ref{class2a}), $\rho^{\phi^\pm}$ and $\rho^{\varphi^\pm}$ are two qubit density matrices corresponding to Bell states. Similar structure as of eq.(\ref{class2a}) can be found if we trace out second qubit (i.e. the second party, which is also a peripheral qubit). Removing third qubit (central qubit) from state (\ref{star}) however makes the $\vert Star\rangle$ state separable. The class of states as defined in eqs.(\ref{class1c}) and (\ref{class2a}) are of NMEMS types. It was also pointed out that the states of the type $p\rho_{sep}+(1-p)\rho_{ent}$, where $\rho_{sep}$ is the two qubit $\vert Star\rangle$ state (after removal of central qubit) and $\rho_{ent}$ is the two qubit Bell state, are not suitable as quantum teleportation channels.\\\\
Though, to define the above convex mixtures we have used separate symbols for the state parameters $p,\:\:q,\:\:r$ and $s$, in the following analysis for the sake of comparative analysis of these states with respect to the study of fidelity of teleportation and Bell violation we have varies any of these state parameters in a defined range.
\section{$X$ type states as teleportation channels and their Bell-CHSH violation:}
\subsection*{$X$ state:}\label{sec:measures1}
Mendon\emph{c}a et al. had shown that there exist two-qubit states which belong to a wider class of states of the form $X$ in which density matrix contains elements only along the main diagonal and the complementary diagonal positions. Such states are called $X$ states \cite{roy2023,nandi2018,rau2009,alim2010,qn2012,mendonca2014}. The density operator structure of such states are given as
\begin{eqnarray}
\label{xstate1}
\mathcal{\rho^{(X)}}=\begin{pmatrix}
\alpha & 0 & 0 & \eta\\
0 & \beta & \xi & 0\\
0 & \xi^{*} & \gamma & 0\\
\eta^{*} & 0 & 0 & \delta\\
\end{pmatrix}.
\end{eqnarray}
The analytic expression for concurrence of this state is given by\\  
\begin{eqnarray}
\label{x-stateconc}
C(\mathcal{\rho^{(X)}})= 2\: \max (0,\mid \xi \mid -\sqrt{\alpha\:\delta},\mid \eta \mid -\sqrt{\beta\:\gamma} ).
\end{eqnarray}
Using eq.(\ref{telep1}), we calculate the teleportation fidelity of this generic $X$ state which is thus obtained as \\
\begin{eqnarray}
\label{telepX}
f^{T}(\mathcal{\rho^{(X)}})=\frac{1}{2}+\frac{1}{6}\sqrt{(2\eta+2\xi)^{2}}+\frac{1}{6}\sqrt{(-2\eta+2\xi)^{2}}+\frac{1}{6}\sqrt{(\alpha-\beta-\gamma+\delta)^{2}}.
\end{eqnarray}
Similarly, for the density matrix of eq.(\ref{xstate1}), the mixedness of $X$ state (using eq.(\ref{le1})) can be calculated as\\
\begin{eqnarray}
\label{mixednessX}
L(\mathcal{\rho^{(X)}})=\frac{4}{3}-\frac{4}{3}(\alpha^{2}+\beta^{2}+\gamma^{2}+\delta^{2})-\frac{8}{3}(\eta^{2}+\xi^{2}).
\end{eqnarray}
Using eq.(\ref{bell-chsh}), we can also immediately identify the eigenvalues of the matrix $T^{\dagger}T$ corresponding to the $X$ state $\mathcal{\rho^{(X)}}$ of eq.(\ref{xstate1}) as 
\begin{eqnarray}
\label{eigenvaluesxstate}
u_{1}&=& 8(\xi^2+\eta^2),\nonumber\\
u_{2} &=& (-2\eta+2\xi)^2+(\alpha - \beta - \gamma + \delta)^2,\nonumber\\
u_{3} &=& (2\eta+2\xi)^2+(\alpha - \beta - \gamma + \delta)^2.
\end{eqnarray}
It is easy to observe that the states of class $1$ defined in eqs.(\ref{class1a}) and (\ref{class1b}) take the structure of $X$ state, while the states represented as class $2$, defined in eqs.(\ref{class1c}) and (\ref{class2a}) are not of the type of $X$ state. We now analyze the concurrence, teleportation fidelity and mixedness of the states having  $X$ state's structure.\\\\
For example, we can consider well-known Werner state \cite{werner1989} which is a maximally entangled mixed state of Ishizaka-Hiroshima form\cite{hiroshima2000}. The Werner state has several forms in $2-$ qubit system, the one which we are interested in, is given by
\begin{eqnarray}
\label{werner state}
\rho^{(werner)} = \frac{1-m}{3}I_{4} + \frac{4m-1}{3}\vert \varphi^{-}\rangle\langle \varphi^{-}\vert,\:\: 0\leq m\leq 1,
\end{eqnarray}
where $\vert \varphi^{-}\rangle$ is the singlet state from eq.(\ref{bell}). Consequently, the density operator form of the state (\ref{werner state}) is
\begin{eqnarray}
\label{werner1}
\rho^{(werner)}=\begin{pmatrix}
\frac{1-m}{3} & 0 & 0 & 0\\
0 & \frac{2m+1}{6} & \frac{1-4m}{6} & 0\\
0 & \frac{1-4m}{6} & \frac{2m+1}{6} & 0\\
0 & 0 & 0 & \frac{1-m}{3}\\
\end{pmatrix}.
\end{eqnarray}
We see that the form of the density matrix of the Werner state of eq.(\ref{werner1}) is of $\mathcal{\rho^{(X)}}$ type. Werner state is a type of state that although is entangled, satisfies Bell-CHSH inequality and that efficacy of Werner state as teleportation channel, is well established \cite{adhikari2010}. Using eq.(\ref{x-stateconc}), the concurrence of the state (\ref{werner1}) is given as
\begin{eqnarray}
\label{concwerner}
C(\rho^{(werner)}) =
\begin{array}{cc}
  \Big\lbrace & 
    \begin{array}{cc}
      \frac{2m+1}{6}, ~~~m > -\frac{1}{2} \\
     \frac{1-2m}{2}, ~~~ m < \frac{1}{2} \\
    \end{array}
\end{array}.
\end{eqnarray}
The state is mixed with mixedness $L(\rho^{(werner)})=\frac{8}{9}-\frac{16}{9}m^{2}+\frac{8}{9}m$ and from eq.(\ref{telepX}), the teleportation fidelity of the Werner state is found as
\begin{eqnarray}
\label{telepwerner}
f^{T}(\rho^{(werner)}) = \frac{1+2m}{3}, 
\end{eqnarray}
where the fidelity exceeds classical limit for $m>\frac{1}{2}$. 
\\\\
We now perform a study of the states defined in sec.\ref{sec:bipartite} on the basis of their utility as teleportation channels. For this, we denote the states defined in eqs.(\ref{class1a})and (\ref{class1b}) as subclass $(A)$ that are of the type of $X$ states and those defined in eqs.(\ref{class1c}) and (\ref{class2a}) as subclass $(B)$ which do not fall into the category of $X$ states. Such states, in this paper, have been termed as non-$X$ states.
\subsection*{Teleportation fidelity and Bell-CHSH violation of Subclass $(A)$:}
In this section we analyze the structure of the states that are of $X$ type. These states have been represented in eqs.(\ref{class1a}) and (\ref{class1b}). 
\subsubsection*{Class $\rho^{(1)}$:}
For the clarity of the discussion we first of all denote the states $p\rho^{W} + (1-p)\rho^{\phi^+}$ of eq.(\ref{class1a}) by $\rho^{(1)_{\phi+}}$ and the states $p\rho^{W} + (1-p)\rho^{\phi^-}$ of eq.(\ref{class1a}) by $\rho^{(1)_{\phi-}}$. When the state is of the form $\rho^{(1)_{\phi+}}$, we see, from eq.(\ref{x-stateconc}) and as concurrence always lies between $0$ and $1$, that when $0.7081< p \leq 1$ the concurrence $C(\rho^{(1)_{\phi+}})$ is $\frac{2}{3}p-2\sqrt{\frac{(3-p)(1-p)}{12}}$, while for $0\leq p < 0.6$, the state's concurrence $C(\rho^{(1)_{\phi+}})$ is $1-p-\frac{2}{3}\sqrt{p^{2}}$. However for all admissible values of the parameter $p$ (see eq.(\ref{class1a})), the state is suitable as teleportation channel and the teleportation fidelity exceeds the classical limit of $\frac{2}{3}$. Using (\ref{mixednessX}), the mixedness $L(\rho^{(1)_{\phi+}})$ of the state is given as $\frac{20}{9}p-\frac{44}{27}p^{2}$. It is easy to observe that when $p=0$, the mixedness is zero which is obvious as the mixture reduces to the Bell state whose purity is $1$. Again when $p=1$, the mixedness is $\frac{16}{27}$, which is the mixedness of $\rho_{AB}^{W}$ \cite{adhikari2010,roy2023}. The teleportation fidelity $f^{T}(\rho^{(1)_{\phi+}})$ of the given state, using eq.(\ref{telepX}), is found to be $\frac{1}{2}+\frac{1}{6}\Big(\sqrt{(1-\frac{p}{3})^{2}}+\sqrt{(-1+\frac{5p}{3})^{2}}+\sqrt{(1-\frac{4p}{3})^{2}}\Big)$. Also for $p=0$, the teleportation fidelity of the given state is $1$, while for $p=1$, the teleportation fidelity of the reduced MEMS $\rho^{W}$ is found to be $\frac{7}{9}$ \cite{adhikari2010,roy2023}. Likewise, by taking the state $\rho^{(1)_{\phi-}}$, we see that the expressions for mixedness $L(\rho^{(1)_{\phi-}})$ and teleportation fidelity $f^{T}(\rho^{(1)_{\phi-}})$ of the state are exactly same as that of $\rho^{(1)_{\phi+}}$. The concurrence  $C(\rho^{(1)_{\phi-}})$ of the state $\rho^{(1)_{\phi-}}$ is however of the form  $\frac{2}{3}p-2\sqrt{\frac{(3-p)(1-p)}{12}}$ and this is valid for $0.7081<p\leq 1$. 
\subsubsection*{Class $\rho^{(2)}$:}
The states $p^{\prime}\rho^{W} + (1-p^{\prime})\varrho^{\varphi^+}$ and $p^{\prime}\rho^{W} + (1-p^{\prime})\rho^{\varphi^-}$ of eq.(\ref{class1a}) are now denoted by $\rho^{(2)_{\varphi+}}$ and $\rho^{(2)_{\varphi-}}$ respectively. Since these states are also of $X$ type, we use eq.(\ref{x-stateconc}), (\ref{telepX}) and (\ref{mixednessX}) to calculate concurrence, teleportation fidelity and mixedness. We find that for $0\leq p^{\prime}\leq 1$, the concurrence $C(\rho^{(2)_{\varphi+}})$ is given by $1-\frac{p^{\prime}}{3}$. The expression for the mixedness $L(\rho^{(2)_{\varphi+}})$ of the state is $p^{\prime}\Big(\frac{8}{9}-\frac{8}{27}p^{\prime}\Big)$. We also see that the state $\rho^{(2)_{\varphi+}}$ is useful in teleportation for $0\leq p^{\prime} \leq 1$ as $N(\rho^{(2)_{\varphi+}})>1$, the teleportation fidelity $f^{T}(\rho^{(2)_{\varphi+}})$ of the given state is found to be $1-\frac{2}{9}p^{\prime}$ and it exceeds classical fidelity $\frac{2}{3}$. As before we also observe that when $p^{\prime} =0$, the state $\rho^{(2)_{\varphi+}}$ reduces to Bell state with teleportation fidelity $1$ while for $p^{\prime} = 1$, the given state reduces to two qubit $\rho^{W}$ with teleportation fidelity $\frac{7}{9}$ and mixedness $\frac{16}{27}$. Again by considering the state $\rho^{(2)_{\varphi-}}$, we see from eq.(\ref{x-stateconc}) that the concurrence $C(\rho^{(2)_{\varphi-}})$ is $1-\frac{p^{\prime}}{3}$ when $0\leq p^{\prime}\leq 1$. Hence we can say the state is entangled for all values of parameter $p^{\prime}$. Also from eqs.(\ref{telepX}) and (\ref{mixednessX}), teleportation fidelity $f^{T}(\rho^{(2)_{\varphi-}})$ and  the mixedness $L(\rho^{(2)_{\varphi-}})$ of the state is found respectively to be $\frac{1}{2}+\frac{\sqrt{(\frac{5p^{\prime}}{3}-1)^{2}}}{3} + \frac{\sqrt{(\frac{2p^{\prime}}{3}-1)^{2}}}{6}$ and $8p^{\prime}\Big(\frac{1}{3}-\frac{7}{27}p^{\prime}\Big)$. It is observed that although the state $\rho^{(2)_{\varphi-}}$ remains entangled for all values of parameter $p^{\prime}$, the state is useful for teleportation with fidelity exceeding classical fidelity of $\frac{2}{3}$ only when $0\leq p^{\prime} <0.5$ or $0.75\leq p^{\prime} \leq 1$. Here also for the values of parameter $p^{\prime}=0$ and $p^{\prime}=1$, the state respectively reduce to Bell state and two qubit $\rho^{W}_{AB}$, in which  the teleportation fidelities are respectively $1$ and $\frac{7}{9}$.
\subsubsection*{Classes $\rho^{(3)}$ and $\rho^{(4)}$:}
It is being further observed that for the states $\rho^{(3)}$ and $\rho^{(4)}$ defined in eq.(\ref{class1b}), the concurrence, teleportation fidelity and mixedness are same as the states $\rho^{(1)}$ and $\rho^{(2)}$ (defined in eq.(\ref{class1a})) respectively.  
\subsubsection*{Comparison of teleportation fidelities of mixtures of Bell states with Subclass $(A)$ and two qubit state derived from $\vert GHZ \rangle$ state:}
The $\vert GHZ\rangle$ state is a three qubit entangled state \cite{green1989}, but the state is different from $\vert W\rangle$ state as when a qubit is lost from $\vert GHZ\rangle$, the reduced two-qubit state is separable whereas under similar conditions $\vert W\rangle$ is inseparable. Hence $\vert GHZ\rangle$ and $\vert W\rangle$ are two different classifications under SLOCC operations in the tripartite scenario. The $\vert GHZ\rangle$ state is defined as \cite{green1990,green1997,cirac2000}
\begin{eqnarray}
\label{ghz}
\vert GHZ\rangle = \frac{1}{\sqrt{2}}(\vert 000\rangle + \vert 111\rangle).
\end{eqnarray}
Now as per our construction of states as defined in the previous sections we eliminate any one qubit from state (\ref{ghz}) to get state $\rho^{ghz}$ and then taking convex combination with any of the four possible Bell states (say $\vert \phi^{+}\rangle$) as 
\begin{eqnarray}
\label{ghzbellmixture}
\rho^{g}&=&  t\rho^{ghz} + (1-t)\rho^{\phi^+}.
\end{eqnarray}
These states are type of $X$ state. It can easily be shown that the above states are useful as quantum teleportation channels and violate Bell-CHSH inequality for $0\leq t\leq 1$. Using eq.(\ref{telepX}) the teleportation fidelity of the state is thus given by
\begin{eqnarray}
\label{telepghz}
f^{T}(\mathcal{\rho}^{g}) = \frac{2}{3} + \frac{1}{3}\Big(1-t\Big).
\end{eqnarray}
We can also construct states like $\rho^{g}$ of eq.(\ref{ghzbellmixture}) by taking convex combinations of the state $\rho^{ghz}$ with other Bell states from eq.(\ref{bell}) and conduct similar study.
\subsection*{Bell-CHSH violation of Subclass $(A)$:}
\subsubsection*{Class $\rho^{(1)}$:}
Using eqs.(\ref{bell-chsh}) and (\ref{eigenvaluesxstate}), we see that for the states of the type $\rho^{(1)_{\phi+}}$, $M(\rho^{(1)_{\phi+}})=4-10p+\frac{67}{9}p^2$. Now $M(\rho^{(1)_{\phi+}})>1$ when $0\leq p\leq 0.45$ and when $0.89\leq p\leq 1$. Hence in these two ranges, the state $\rho^{(1)_{\phi+}}$ violates Bell-inequality. But when $0.45<p<0.89$, the state $\rho^{(1)_{\phi+}}$ satisfies Bell's inequality, although the state is entangled there. Again for the state $\rho^{(1)_{\phi-}}$ , $M(\rho^{(1)_{\phi-}})=4-\frac{22}{3}p+\frac{43}{9}p^2$. It is being observed that for the state $\rho^{(1)_{\phi-}}$, $M(\rho^{(1)_{\phi-}})>1$ for all values of the state parameter $p$. Therefore, the state $\rho^{(1)_{\phi-}}$, violates Bell's inequality $\forall$ values of parameter $p$. Since the concurrence of the state $\rho^{(1)_{\phi+}}$ is positive in the range $[0,0.6]$ and in $(0.7,1]$ and as per the nature of Bell violation by the state $\rho^{(1)_{\phi+}}$, as discussed, it is thereby observed that the states of the form $\rho^{(1)_{\phi+}}$ satisfies Bell's inequality although being entangled in the ranges $0.45\leq p <0.6$ and for $0.70<p\leq 0.89$. The state $\rho^{(1)_{\phi-}}$ is having positive concurrence only when $0.70<p\leq 1$ and in this range it violates Bell's inequality too.\\\\
Thus we see that the class of states of the form $\rho^{(1)}$ is useful for teleportation although for certain ranges of the state parameter $p$, $\rho^{(1)}$ satisfies Bell's inequality although being entangled in the given range.
\subsubsection*{Class $\rho^{(2)}$:} 
As before using eqs.(\ref{bell-chsh}) and (\ref{eigenvaluesxstate}), we see that for the states of the type $\rho^{(2)_{\varphi+}}$,  $M(\rho^{(2)_{\varphi+}}) =\frac{7}{9}(p^{\prime})^{2}-\frac{10}{3}p^{\prime}+4$. Now $M(\rho^{(2)_{\varphi+}})>1$, $\forall$ $p^{\prime}$. On the other hand, for the states $\rho^{(2)_{\varphi-}}$, $M(\rho^{(2)_{\varphi-}})=\frac{79}{9}(p^{\prime})^{2}-\frac{34}{2}p^{\prime}+4$ which is greater than unity when $0\leq p^{\prime}<0.37$ and when $0.91< p^{\prime}\leq 1$. For $p^{\prime}\in [0.37,\:0.91]$, the state $\rho^{(2)_{\varphi-}}$ satisfies Bell's inequality. Now both the types of the states $\rho^{(2)_{\varphi+}}$ and $\rho^{(2)_{\varphi-}}$ are entangled for $0\leq p^{\prime}\leq 1$, but for certain ranges of the parameter $p^{\prime}$, the states of the type $\rho^{(2)}$ satisfy Bell's inequality.\\\\
Thus it is being observed that, in the two regions of the parameter $p^{\prime}$ such as $[0,\:0.5)$ and $[0.75,\:1]$, the states $\rho^{(2)}$ is entangled, satisfies Bell's inequality and moreover are useful as quantum teleportation channels.\\\\
We now plot the teleportation fidelities of the states (\ref{class1a}), (\ref{class1b}) and (\ref{ghzbellmixture}) (although for explicit representation of the states we have used different symbols for state parameters, these state parameters all vary between $0$ and $1$ and hence to compare we have varied one such state parameter $p$).
\begin{figure}[h]
\label{fig1}
\includegraphics[width=14.14cm]{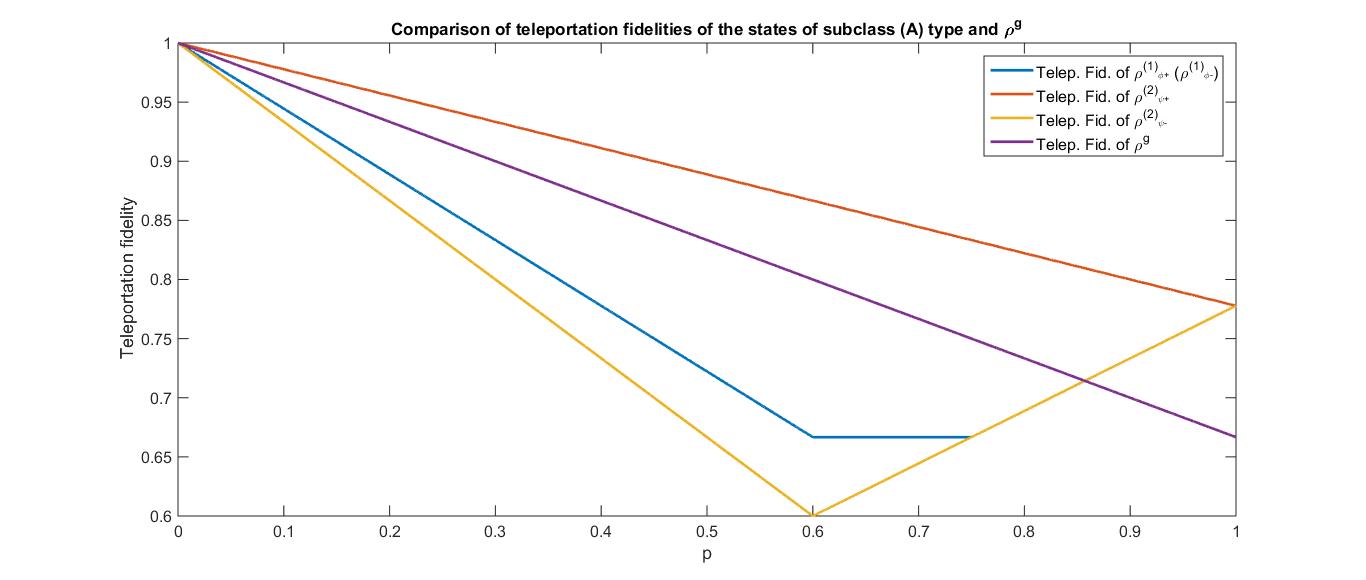}
\caption{The figure shows the plots of teleportation fidelities of the states $\rho^{(1)}$ (or $\rho^{(3)}$) (blue line), $\rho^{(2)}$ (or $\rho^{(4)}$) (red line) and $\rho^{g}$ (yellow line) defined respectively in eqs.(\ref{class1a}),(\ref{class1b}) and (\ref{ghzbellmixture}) against the parameter $p$ (the other symbols $q$ and $t$ used to defined the respective mixtures are similar to $p$ and ranges from $0$ to $1$).}
\end{figure}\\\\
The figure $1$ shows the plots of the teleportation fidelity $\rho^{(1)_{\phi+}}$ (and $\rho^{(1)_{\phi-}}$), $\rho^{(1)_{\psi-}}$, $\rho^{(1)_{\psi+}}$ and $\rho^{g}$. The state parameters represented in eqs.(\ref{class1a}), (\ref{class1b}) and (\ref{ghz}) are $p,\:p^{\prime},\:t$ but as all these state parameters vary from $0$ to $1$. For the sake of discussion we have plotted the teleportation fidelities of these states with respect to a common framework and hence we have arbitrarily taken parameter $p$ along horizontal axis varying from $0$ to $1$. We see from fig.$1$ that the teleportation fidelity of $\rho^{(1)_{\phi+}}$ (and $\rho^{(1)_{\phi-}}$) (the expression for both of these states are similar) as well as that of $\rho^{(1)_{\psi-}}$ are exceeding the teleportation fidelity of $\rho^{g}$ when $0.85\leq p \leq 1$ while the teleportation fidelity of  $\rho^{(1)_{\psi+}}$ is always exceeding that of $\rho^{g}$ for $0\leq p\leq 1$. This implies that the $X$ type states $\rho^{(1)}$ and $\rho^{(2)}$ described in eqs.(\ref{class1a}) and (\ref{class1b}) give better performance as quantum teleportation channels than the state $\rho^{g}$ described in eq.(\ref{ghz}) for certain ranges of states' parameter whereas for some other range the state $\rho^{g}$ described in eq.(\ref{ghz}) is better performer as teleportation channel. The analysis of fig.$1$ has been summarized in the following table.
\begin{table}[h!]
\begin{center}
\caption{Summary of teleportation fidelities of the states $\rho^{(1)}$, $\rho^{(2)}$ and $\rho^{g}$}
\label{table1}
\begin{tabular}{|c|c|c|c|c|}
\hline
$p$ & Telep. Fid. of $\rho^{(1)_{\phi+}}$ (or $\rho^{(1)_{\phi-}}$) & Telep. Fid. of $\rho^{(2)_{\psi+}}$ & Telep. Fid. of $\rho^{(2)_{\psi-}}$ & Telep. Fid. of $\rho^{g}$ \\
\hline
$0.0$ & $1$ & $1$ & $1$ & $1$\\
\hline
$0.1$ & $0.94$ & $0.98$ & $0.93$ & $0.97$\\
\hline
$0.2$ & $0.89$ & $0.96$ & $0.87$ & $0.93$\\
\hline
$0.3$ & $0.83$ & $0.93$ & $0.80$ & $0.90$\\
\hline
$0.4$ & $0.78$ & $0.91$ & $0.73$ & $0.87$\\
\hline
$0.5$ & $0.72$ & $0.89$ & $0.67$ & $0.83$\\
\hline
$0.6$ & $0.67$ & $0.87$ & $0.60$ & $0.80$\\
\hline
$0.7$ & $0.67$ & $0.84$ & $0.64$ & $0.77$\\
\hline
$0.8$ & $0.69$ & $0.82$ & $0.69$ & $0.73$\\
\hline
$0.9$ & $0.73$ & $0.80$ & $0.73$ & $0.70$\\
\hline
$1$ & $0.78$ & $0.78$ & $0.78$ & $0.67$\\
\hline
\end{tabular}
\end{center}
\end{table}
\section{Non-$X$ type states as teleportation channels and their Bell-CHSH violation:}\label{sec:nonX}
The states whose structures are not similar to that of $X$ state as defined in eq.(\ref{xstate1}), are termed in this paper as non-$X$ states.
\subsection*{Teleportation fidelity and Bell-CHSH violation of Subclass $(B)$:}
To begin with the analysis of the states of subclass $(B)$, we observe that the convex combination of the two qubit state derived from $\vert W\bar{W}\rangle$ with $\phi^\pm$ of class of states $\rho^{(5)}$ defined in eq.(\ref{class1c}) can be represented by the following general density matrix form.
\begin{eqnarray}
\label{nonxstate1}
\mathcal{\rho^{C}}^{1}=\begin{pmatrix}
\alpha +\delta & \alpha & \alpha & \delta\\
\alpha & 2\alpha & 2\alpha & \alpha\\
\alpha & 2\alpha & 2\alpha & \alpha\\
\delta & \alpha & \alpha & \alpha +\delta\\
\end{pmatrix}
\end{eqnarray}
Using eqs. (\ref{conc1}), (\ref{telep1}) and (\ref{le1}) we calculate the concurrence, teleportation fidelity and mixedness of the state. The concurrence of the state is thus denoted by $C(\mathcal{\rho^{C}}^{1})$ and is given by
\begin{eqnarray}
\label{conclass1c}
C(\mathcal{\rho^{C}}^{1}) = \sqrt{d_{1}+\frac{\sqrt{d_{2}}}{2}} - \sqrt{d_{1}-\frac{\sqrt{d_{2}}}{2}} - \sqrt{d_{3}},
\end{eqnarray}
where $d_{1} = \frac{9}{2}\alpha^{2}+2\alpha\delta+2\delta^{2}$,$d_{2}=81\alpha^{4}+72\alpha^{3}\delta-168\alpha^{2}\delta^{2}+32\alpha\delta^{3}+16\delta^{4}$ and $d_{3}=\alpha^{2}$.\\\\ Now the eigenvalues of of the matrix $T^{\dagger}T$ are
\begin{eqnarray}
\label{eigenvaluesrhoc1}
v_{1} &=& (2\alpha - 2\delta)^2,\nonumber\\
v_{2} &=& (4\alpha + 2\delta )^2,\nonumber\\
v_{3} &=& (4\alpha - 2\delta)^2,
\end{eqnarray}
and consequently the teleportation fidelity of the given state (\ref{nonxstate1}) is denoted by $f^{T}(\mathcal{\rho^{C}}^{1})$ and is found as
\begin{eqnarray}
\label{telepclass1c}
f^{T}(\mathcal{\rho^{C}}^{1}) = \frac{1}{2}+\frac{1}{6}\sqrt{(2\alpha - 2\delta)^{2}}+\frac{1}{6}\sqrt{(2\delta +4\alpha)^{2}}+\frac{1}{6}\sqrt{(-2\delta +4\alpha)^{2}},
\end{eqnarray}
while the mixedness of state (\ref{nonxstate1}) is denoted by $L(\mathcal{\rho^{C}}^{1})$ and is obtained as
\begin{eqnarray}
\label{mixedclass1c}
L(\mathcal{\rho^{C}}^{1}) = \frac{4}{3}-\frac{8}{3}(\alpha + \delta)^{2}-32\alpha^{2}-\frac{8}{3}\delta^{2}.
\end{eqnarray}
Again the states $\rho^{(6)}$ defined in eq.(\ref{class1c}) can be categorized in two different forms, which we respectively denote by $\rho^{(6a)} = r^{\prime}\rho^{W\bar{W}} + (1-r^{\prime})\vert \varphi^{+}\rangle\langle \varphi^{+}\vert$ and $\rho^{(6b)} = r^{\prime}\rho^{W\bar{W}} + (1-r^{\prime})\vert \varphi^{-}\rangle\langle \varphi^{-}\vert$. The density matrices of $\rho^{(6a)}$ and $\rho^{(6b)}$ are respectively denoted by $\mathcal{\rho^{C}}^{2a}$ and $\mathcal{\rho^{C}}^{2b}$ and is found of the following forms.
\begin{eqnarray}
\label{nonxstate2}
\mathcal{\rho^{C}}^{2a}=\begin{pmatrix}
\alpha  & \alpha & \alpha & 0\\
\alpha & 2\alpha+ \beta & 2\alpha+ \beta & \alpha\\
\alpha & 2\alpha+ \beta & 2\alpha+ \beta & \alpha\\
0 & \alpha & \alpha & \alpha\\
\end{pmatrix},\:\:
\mathcal{\rho^{C}}^{2b}=\begin{pmatrix}
\alpha  & \alpha & \alpha & 0\\
\alpha & 2\alpha+ \beta & 2\alpha- \beta & \alpha\\
\alpha & 2\alpha- \beta & 2\alpha+ \beta & \alpha\\
0 & \alpha & \alpha & \alpha\\
\end{pmatrix}.
\end{eqnarray}
As before we use eqs. (\ref{conc1}), (\ref{telep1}) and (\ref{le1}) to calculate the concurrence, teleportation fidelity and mixedness of the states defined in eq.(\ref{nonxstate2}).\\\\
The concurrence of $\mathcal{\rho^{C}}^{2a}$ is given by
\begin{eqnarray}
\label{conc2a}
C(\mathcal{\rho^{C}}^{2a}) = \sqrt{f_{1}+\frac{\sqrt{f_{2}}}{2}} - \sqrt{f_{1}-\frac{\sqrt{f_{2}}}{2}} - \sqrt{f_{3}},
\end{eqnarray}
where  $f_{1} = \frac{9}{2}\alpha^{2}+8\alpha\beta+2\beta^{2}$,$f_{2}=81\alpha^{4}+288\alpha^{3}\beta+312\alpha^{2}\beta^{2}+128\alpha\beta^{3}+16\beta^{4}$ and $f_{3}=\alpha^{2}$.\\\\
The eigenvalues of the matrix $T^{\dagger}T$ are 
\begin{eqnarray}
\label{eigenvaluesrho21}
w_{1}&=& (4\alpha + 2\beta)^2,\nonumber\\
w_{2}&=& (4\alpha + 2\beta)^2,\nonumber\\
w_{3}&=& (-2\alpha - 2\beta)^2.
\end{eqnarray}
so that the teleportation fidelity of the state $\mathcal{\rho^{C}}^{2a}$ is  found to be
\begin{eqnarray}
\label{telep2a}
f^{T}(\mathcal{\rho^{C}}^{2a}) = \frac{1}{2} +\frac{1}{3}\sqrt{(4\alpha+2\beta)^{2}}+\frac{1}{6}\sqrt{(-2\alpha-2\beta)^{2}},
\end{eqnarray}
while using eq.(\ref{le1}), the mixedness of the state $\mathcal{\rho^{C}}^{2a}$ is
\begin{eqnarray}
\label{mixed2a}
L(\mathcal{\rho^{C}}^{2a}) = \frac{4}{3} -\frac{40}{3}\alpha^{2}-\frac{16}{3}(2\alpha+ \beta)^{2}.
\end{eqnarray}
Again using eqs.(\ref{conc1}),(\ref{telep1}) and (\ref{le1}) we calculate the concurrence, teleportation fidelity and mixedness of the state $\mathcal{\rho^{C}}^{2b}$. The concurrence of $\mathcal{\rho^{C}}^{2b}$ is given by
\begin{eqnarray}
\label{conc2b}
C(\mathcal{\rho^{C}}^{2b}) = 2\sqrt{\alpha^2}-2\sqrt{\beta^2}.
\end{eqnarray}
The eigenvalues of the matrix $T^{\dagger}T$ are 
\begin{eqnarray}
\label{eigenvaluesrho22}
z_{1}&=& (4\alpha - 2\beta)^2,\nonumber\\
z_{2}&=& (4\alpha - 2\beta)^2,\nonumber\\
z_{3}&=& (-2\alpha - 2\beta)^2.
\end{eqnarray}
The teleportation fidelity and the mixedness of the state $\mathcal{\rho^{C}}^{2b}$ are however obtained as
\begin{eqnarray}
\label{mixed2b}
f^{T}(\mathcal{\rho^{C}}^{2b})&=&\frac{1}{2}+\frac{\sqrt{(4\alpha-2\beta)^2}}{3}+\frac{\sqrt{(-2\alpha-2\beta)^2}}{6},\nonumber\\
L(\mathcal{\rho^{C}}^{2b}) &=& \frac{4}{3}-\frac{104\alpha^2}{3}-\frac{16\beta^2}{3}.
\end{eqnarray}
\subsubsection*{Teleportation fidelity of class $\rho^{(5)}$:}
First of all, the convex combination of $\rho^{W\bar{W}}$ and $\phi^+$ is denoted by $\mathcal{\rho^{C}}^{1}_{\phi^+}$ and the the convex combination of $\rho^{W\bar{W}}$ and $\phi^-$ is denoted by $\mathcal{\rho^{C}}^{1}_{\phi^-}$.  Using eqs.(\ref{conclass1c}), we observe that the concurrence of state $\mathcal{\rho^{C}}^{1}_{\phi^+}$ are respectively given as $C(\mathcal{\rho^{C}}^{1}_{\phi^+}) = \frac{\sqrt{u+6\sqrt{v}}}{12}-\frac{\sqrt{u-6\sqrt{v}}}{12}-\frac{\sqrt{r^2}}{6}$, where $u=18r^2+48r\frac{1-r}{2}+288\frac{(1-r)^2}{4}$ and $v=-135r^4+72r^3+408r^2-480r+144$. The concurrence is positive when $0\leq r<0.6$ and when $0.75<r\leq 1$. From eqs.(\ref{telepclass1c}) and (\ref{mixedclass1c}) we find the teleportation fidelity and mixedness of the state $\mathcal{\rho^{C}}^{1}_{\phi^+}$ respectively as $f^{T}(\mathcal{\rho^{C}}^{1}_{\phi^+})=\frac{1}{2}+\frac{1}{6}[\sqrt{(\frac{4r}{3}-1)^2}+\sqrt{(1-\frac{r}{3})^2}+\sqrt{(-1+\frac{5r}{3})^2}]$ and $L(\mathcal{\rho^{C}}^{1}_{\phi^+})=-\frac{50}{27}r^2+\frac{20}{9}r$ while the state is mixed for all $r$. For parameter $0\leq r \leq 1$, the state $\mathcal{\rho^{C}}^{1}_{\phi^+}$ of eq.(\ref{class1c}) $N(\mathcal{\rho^{C}}^{1}_{\phi^+})>1$ and teleportation fidelity of the state exceeding the classical fidelity of $\frac{2}{3}$. But the state $\mathcal{\rho^{C}}^{1}_{\phi^+}$ is entangled for $r\in[0,\:0.6)$ and $(0.75,\:1]$. So, in this two ranges the state can bes suitably used as quantum teleportation channel successfully. It is also to be noted that at the extreme points i.e. when $r=0$ or $r=1$, the state's concurrence, teleportation fidelity and mixedness are aligned with that of Bell state and two qubit $\rho^{W\bar{W}}$ from eq.(\ref{class1c}). Similarly, for the state $\mathcal{\rho^{C}}^{1}_{\phi^-}$, using eq.(\ref{conclass1c}), it can be easily shown that the state $\mathcal{\rho^{C}}^{1}_{\phi^-}$ is entangled when $0\leq r\leq 0.39$ and when $0.87<r\leq 1$. Also, $N(\mathcal{\rho^{C}}^{1}_{\phi^-})>1$ and $f^{T}(\mathcal{\rho^{C}}^{1}_{\phi^-})>\frac{2}{3}$ for all $r$. Hence the state $\mathcal{\rho^{C}}^{1}_{\phi^-}$ can suitably used as quantum teleportation channel for $0\leq r\leq 0.39$ and when $0.87<r\leq 1$.
\subsubsection*{Teleportation fidelity of class $\rho^{(6)}$:}
The concurrence of state $\mathcal{\rho^{C}}^{2a}$ of (\ref{nonxstate2}) has concurrence of the form $C(\mathcal{\rho^{C}}^{2a})$ which is $\frac{x+6\sqrt{y}}{12}-\frac{x-6\sqrt{y}}{12}-\frac{\sqrt{r^{\prime^{2}}}}{6}$, where $x = 18r^{\prime^{2}}+192r^{\prime}\frac{1-r^{\prime}}{2}+288\frac{(1-r^{\prime})^2}{2}$ and $y = -15r^{\prime^{4}}+48r^{\prime^{3}}+24r^{\prime^{2}}-192r+144$. The mixedness of the state is $L(\mathcal{\rho^{C}}^{2a})=\frac{8}{9}r^{\prime}-\frac{14}{27}r^{\prime^{2}}$ and the teleportation fidelity $f^{T}(\mathcal{\rho^{C}}^{2a})$ is $\frac{1}{2}+\frac{1}{3}\sqrt{(1-\frac{r^{\prime}}{3})^{2}}+\frac{\sqrt{(\frac{2r^{\prime}}{3}-1)^{2}}}{6}$. It is being observed that the state $\mathcal{\rho^{C}}^{2a}$ is entangled with non-negative concurrence $0\leq r^{\prime}\leq 1$. Also $N(\mathcal{\rho^{C}}^{2a})>1$ and $f^{T}(\mathcal{\rho^{C}}^{2a})>\frac{2}{3}$ for all $r^{\prime}$.\\\\
Again for the state $\mathcal{\rho^{C}}^{2b}$ of eq.(\ref{class1c}), the teleportation fidelity, concurrence and mixedness as obtained by eqs.(\ref{conc2b})and (\ref{mixed2b}), are $C(\mathcal{\rho^{C}}^{2b})=\frac{\sqrt{r^{\prime^{2}}}}{3}-2\sqrt{(\frac{1-r^{\prime}}{2})^2}$, $f^{T}(\mathcal{\rho^{C}}^{2b})=\frac{1}{2}+\frac{\sqrt{(\frac{2r^{\prime}}{3}-1)^{2}}}{6}+\frac{\sqrt{(\frac{5r^{\prime}}{3}-1)^{2}}}{3}$, and $L(\mathcal{\rho^{C}}^{2b})=\frac{8}{3}r^{\prime}-\frac{62}{27}r^{\prime^{2}}$. It is seen that the state $\mathcal{\rho^{C}}^{2b}$ is entangled with non-negative concurrence when $0.75\leq r^{\prime}\leq 1$. The state is useful as quantum teleportation channel when $0\leq r^{\prime}\leq 0.5$ and $0.75\leq r^{\prime}\leq 1$ as in these ranges $N(\mathcal{\rho^{C}}^{2b})>1$ and teleportation fidelity exceeds classical limit of $\frac{2}{3}$.
\subsubsection*{Bell-CHSH violation of class $\rho^{(5)}$:}
For the class of states $\mathcal{\rho^{C}}^{1}_{\phi^+}$ of eq.(\ref{class1c}), we observed that the concurrence is greater than zero $\forall$ $r$. Also using eqs.(\ref{bell-chsh}) and (\ref{eigenvaluesrhoc1}) we calculate $M(\mathcal{\rho^{C}}^{1}_{\phi^+})$ which is $(\frac{4r}{3}-1)^2+(1-\frac{r}{3})^2$. Now $M(\mathcal{\rho^{C}}^{1}_{\phi^+})>1$ when $0\leq r\leq 0.38$ whereas for $0.38<r\leq 1$ satisfies Bell's inequality. Therefore the given state is entangled violating Bell's inequality for $0\leq r\leq 0.38$ while for $0.38<r<0.6$, the state is entangled satisfying Bell's inequality. The given state is also entangled satisfying Bell's inequality when $0.75<r\leq 1$. Again, $M(\mathcal{\rho^{C}}^{1}_{\phi^-})=(1-\frac{r}{3})^2+(-1+\frac{5r}{3})^2$ and it is greater than unity when $0\leq r\leq 0.32$ whereas $M(\mathcal{\rho^{C}}^{1}_{\phi^-})<1$ when  $0.32< r\leq 1$. Combining the facts, we can immediately conclude that the state $\mathcal{\rho^{C}}^{1}_{\phi^-}$ is entangled violating Bell's inequality for $r\in [0,\:0.32]$ and is entangled satisfying Bell's inequality for $r\in (0.32,0.39)$ and also for $r\in (0.87,\:1]$.\\\\
Thus we see that the state $\mathcal{\rho^{C}}^{1}_{\phi^+}$, although satisfies Bell's inequality in the ranges $(0.38,\:0.6)$ and $(0.75,\:1]$, is useful as quantum teleportation channel. Also the state   $\mathcal{\rho^{C}}^{1}_{\phi^-}$ is useful as quantum teleportation channel although the state satisfies Bell's inequality in the ranges $0.32<r<0.38$ and $0.87<r\leq 1$. Hence is is seen that the class of states $\rho^{(5)}$ is another example of states which can be suitably used as teleportation channel while satisfying Bell's inequality for some specified ranges of the state parameter $r$.
\subsubsection*{Bell-CHSH violation of class $\rho^{(6)}$:}
With respect to the states defined by $\rho^{(6)}$ of eq.(\ref{class1c}), next we observe that the state $\mathcal{\rho^{C}}^{2a}$ of eq.(\ref{nonxstate2}) is entangled and useful for the purpose of teleportation in the entire admissible range of parameter $r^{\prime}$ in spite of satisfying Bell-inequality in the range $0.878 \leq r^{\prime}\leq 1$. Also we found that the state $\mathcal{\rho^{C}}^{2b}$ represented in eq.(\ref{nonxstate2}) which also corresponds to the state $\rho^{(6)}$ of eq.(\ref{class1c}), satisfies Bell-CHSH inequality when $0.87<r\leq 1$.\\\\
This again leads to the conclusion that the states defined by $\rho^{(6)}$ can be useful as teleportation channel although such states satisfy Bell-CHSH inequality at certain regions of the states' parameter.
\subsubsection*{Comparison of teleportation fidelities of NMEMS with MEMS:}
The teleportation fidelities of the states $\rho^{W}$ and $\rho^{\bar{W}}$ of eq.(\ref{mems1}) are $\frac{7}{9}$ and these states are Ishizaka-Hiroshima class of maximally entangled mixed states. As compared to these MEMS,  the states $\rho^{(1)}$ (or $\rho^{(2)}$), $\rho^{(3)}$ (or $\rho^{(4)}$) ($X$ type NMEMS of eqs.(\ref{class1a}) and (\ref{class1b})) and $\rho^{(5)}$ (or $\rho^{(6)}$) (non-$X$ type NEMS of eq.(\ref{class1c})) give better performance as quantum teleportation channels for certain range of state parameters. The following figure shows this comparative assessment of teleportation fidelities of these states.\\
\begin{figure}[h]
\label{fig2}
\includegraphics[width=16.20cm]{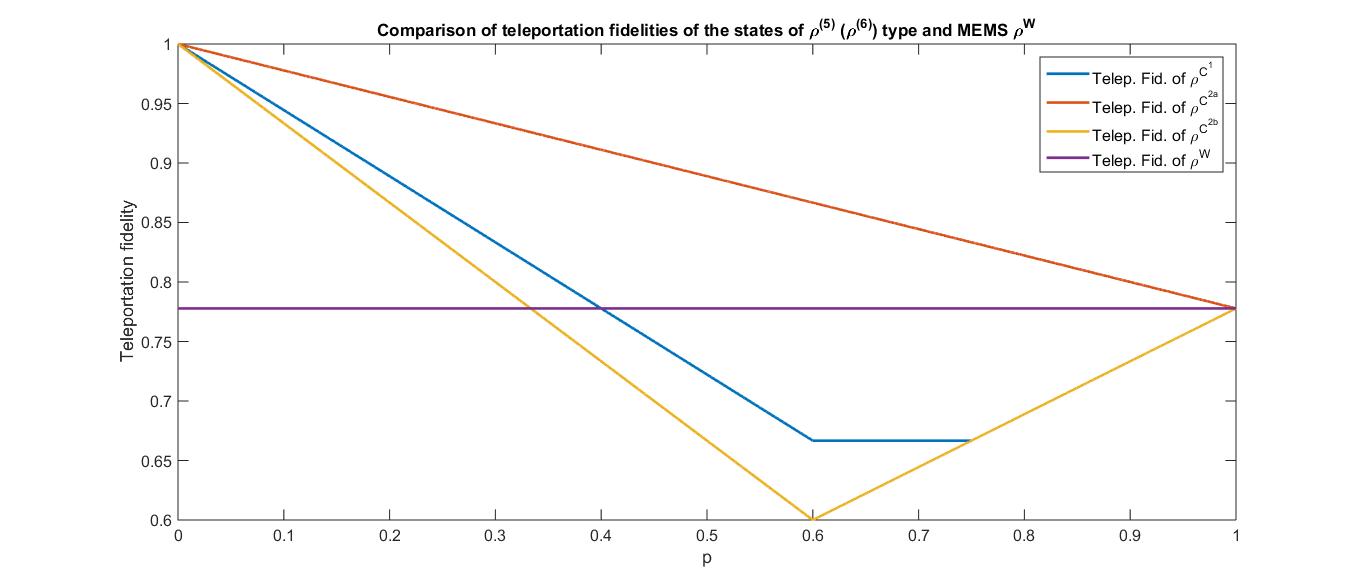}
\caption{The figure shows the plots of teleportation fidelity of the MEMS i.e. $f^{T}(\rho^{W})$ (or of $f^{T}(\rho^{\bar{W}})$) with value $\frac{7}{9}$ for $0\leq p\leq 1$ (line parallel to horizontal axis) and the teleportation fidelities of the other NMEMSs $\rho^{(5)}$ and $\rho^{(6)}$.}
\label{fig1}
\end{figure}\\
It is to be noted that $\rho^{(5)}$ states of eq.(\ref{class1c}) whose form is of non-$X$ type has the teleportation fidelity similar to that of $X$ type states $\rho^{(1)}$ and $\rho^{(3)}$ of eqs.(\ref{class1a}) and (\ref{class1b}). $\rho^{(6)}$ has been described by two forms viz. $\rho^{C^{2a}}$ and $\rho^{C^{2b}}$ of eq.(\ref{nonxstate2}). They are non-maximally entangled mixed states of non-$X$ type. The teleportation fidelity of $\rho^{C^{2a}}$ is much higher than $\rho^{C^{2b}}$. The horizontal line in the fig.$2$ represents the teleportation fidelity of $\rho^{W}$ (and $\rho^{\bar{W}}$) of eq.(\ref{mems1}) which is $\frac{7}{9}$ for both of these MEMS. It is being observed that $\rho^{C^{1}}$ (which are basically of $\rho^{(5)}$ types) performs better as quantum teleportation channel than $\rho^{C^{2b}}$ (which are of $\rho^{(6)}$ type). Also for $0\leq p\leq 0.35$ the teleportation fidelity of $\rho^{C^{2b}}$ is higher than that of $\rho^{W}$ while for $0\leq p\leq 0.4$, the teleportation fidelity of $\rho^{C^{1}}$ is greater than that of $\rho^{W}$. Also it is interesting to note down that the states $\rho^{C^{2a}}$ acts as better candidate as quantum teleportation channel than the maximally entangled mixed state for $0\leq p\leq 1$. (It is to be remembered that in this comparative study too, we have plotted all the teleportation fidelities against a common state parameter, say $p$).
\subsection{Teleportation fidelity and Bell-CHSH violation of mixture of two qubit state derived from $\vert Star\rangle$ and Bell-states:}
We here consider the state $\tau^{(1)}$ defined in eq.(\ref{class2a}) and study the teleportation fidelitiy of the states along with its behaviour with respect to Bell-CHSH violation. We also observe that the states $\tau^{(1)}$ fall into the class of non $X$ type. For the clarity of the inspection we define the mixture of $\rho^{star}$ and Bell-state $\vert \phi^{+}\rangle$ from state $\tau^{(1)}$ as $\tau^{\phi+}$ and the mixture of $\rho^{star}$ and Bell-state $\vert \phi^{-}\rangle$ from state $\tau^{(1)}$ as $\tau^{\phi-}$. The density matrices of the non-$X$ type classes to which they belong are given as follows.
\begin{eqnarray}
\label{nonxstate3}
\mathcal{\tau^{\phi+}}=\begin{pmatrix}
2\alpha  & \alpha-\beta & 0 & \alpha+\beta\\
\alpha-\beta & \alpha-\beta  & 0 & \alpha-\beta\\
0 & 0 & 0 & 0\\
\alpha+\beta & \alpha-\beta & 0 & \alpha+\beta\\
\end{pmatrix},\:\:
\mathcal{\tau^{\phi-}}=\begin{pmatrix}
2\alpha  & \alpha-\beta & 0 & \alpha-3\beta\\
\alpha-\beta & \alpha-\beta  & 0 & \alpha-\beta\\
0 & 0 & 0 & 0\\
\alpha-3\beta & \alpha-\beta & 0 & \alpha+\beta\\
\end{pmatrix}.
\end{eqnarray}
Using eqs.(\ref{conc1}),(\ref{telep1}) and (\ref{le1}) we determine the concurrence, teleportation fidelity and mixedness of the state (\ref{nonxstate3}) which are summarized below. Thus the concurrence, teleportation fidelity and mixedness of the states are denoted respectively by $C(\tau^{\phi\pm})$, $f^{T}(\tau^{\phi\pm})$ and $L(\tau^{\phi\pm})$ and are given by,
\begin{eqnarray}
\label{concnonxstate3}
C(\tau^{\phi+}) &=& \sqrt{\alpha+\beta}\Big(\sqrt{3\alpha+\beta+2\sqrt{2\alpha(\alpha+\beta)}}-\sqrt{3\alpha+\beta-2\sqrt{2\alpha(\alpha+\beta)}}\Big),\nonumber\\
C(\tau^{\phi-}) &=& \sqrt{u+2\sqrt{v}}-\sqrt{u-2\sqrt{v}} ,
\end{eqnarray}
where  $u = 3\alpha^{2}-4\alpha\beta+9\beta^{2}$ and $v=2\alpha^{4}-10\alpha^{3}\beta+6(\alpha\beta)^{2}+18\alpha\beta^{3}$.
\begin{eqnarray}
\label{telepnonxstate3}
f^{T}(\tau^{\phi+}) &=& \frac{1}{2}+\frac{1}{3}\sqrt{(\alpha+\beta)^{2}}+\frac{2}{3}\sqrt{2(\alpha^{2}+\beta^{2})},\nonumber\\
f^{T}(\tau^{\phi-}) &=& \frac{1}{2}+\frac{\sqrt{(\alpha +\beta)^{2}}}{3}+\frac{1}{3}\Big(\sqrt{u^{\prime}+v^{\prime}}+\sqrt{u^{\prime}-v^{\prime}}\Big),
\end{eqnarray}
where $u^{\prime}=2\alpha^{2}-4\alpha\beta+6\beta^{2}$ and $v^{\prime}=4\sqrt{2}(\beta^{2}-\alpha\beta)$.\\\\
and
\begin{eqnarray}
\label{mixednonxstate3}
L(\tau^{\phi+}) &=& \frac{4}{3}-16\alpha^{2}+\frac{16}{3}(\alpha\beta-2\beta^{2}),\nonumber\\
L(\tau^{\phi-}) &=& \frac{4}{3}-16\alpha^{2}+\frac{80}{3}\alpha\beta-32\beta^{2}.
\end{eqnarray}
\subsubsection*{Teleportation fidelities of $\mathcal{\tau^{\phi\pm}}$ and $\mathcal{\tau^{\psi\pm}}$:}
Using eqs.(\ref{concnonxstate3}),(\ref{telepnonxstate3}) and (\ref{mixednonxstate3}), it is being observed that for the state $\tau^{\phi+}$, the concurrence, teleportation fidelity and mixedness are given to be $C(\tau^{\phi+})=\sqrt{\frac{1+s}{4}}\sqrt{\frac{3}{4}+\frac{s}{4}+\frac{1}{2}\sqrt{2+2s}}-\sqrt{\frac{3}{4}+\frac{s}{4}-\frac{1}{2}\sqrt{2+2s}}$, $f^{T}(\tau^{\phi+})=\frac{1}{2}+\frac{1}{3}\sqrt{\frac{1}{4}(1+s)^{2}}+\frac{1}{6}\sqrt{2s^{2}+2}$ and $L(\tau^{\phi+})=\frac{1}{3}(1+s-2s^{2})$. It is found that the state $\tau^{\phi+}$ is entangled for $0\leq s\leq 1$ and can be used as quantum teleportation channel as $N(\tau^{\phi+})>1$ in this range, teleportation fidelity being greater than $\frac{2}{3}$ also. Moreover, for the state $\tau^{\phi-}$, the the concurrence, teleportation fidelity and mixedness are given to be $C(\tau^{\phi-})=\frac{1}{4}\Big(\sqrt{3-4s+9s^{2}+2\sqrt{18s^{3}+6s^{2}-10s+2}}-\sqrt{3-4s+9s^{2}-2\sqrt{18s^{3}+6s^{2}-10s+2}}\Big)$, $f^{T}(\tau^{\phi-})=\frac{1}{2}+\frac{1}{3}\sqrt{(\frac{3s+1}{4})^{2}}+\frac{1}{12}\Big(\sqrt{4\sqrt{2}(s^{2}-s)+2-4s+6s^{2}}+\sqrt{4\sqrt{2}(s-s^{2})+2-4s+6s^{2}}\Big)$ and $L(\tau^{\phi-})=\frac{1}{3}(1+5s)-2s^{2}$. Just like the state $\tau^{\phi+}$, the state $\tau^{\phi-}$ is also entangled for $0\leq s\leq 1$ and the state is mixed in this region. The state can be used as teleportation channel as $N(\tau^{\phi-})>1$ when $0\leq s\leq 1$ with teleportation fidelity $f^{T}(\tau^{\phi-})$ exceeding $\frac{2}{3}$ in two different regions, one when $0\leq s\leq 0.314$ and another when $0.43\leq s\leq 1$.
It is also to be noted that when $s=0$ state $\tau^{(1)}$ is maximally entangled pure state ($\vert \phi^{+}\rangle$ or $\vert \phi^{-}\rangle$) which is a pure state of course with unit teleportation fidelity whereas when $s=1$, the state is $\rho^{star}$ which is a two qubit state derive from $3-$ qubit $\vert Star\rangle$ state with concurrence $\frac{1}{2}$ and teleportation fidelity as $0.81$.
\subsubsection*{Bell-CHSH violation of $\mathcal{\tau^{\phi\pm}}$ and $\mathcal{\tau^{\psi\pm}}$:}
Using eq.(\ref{bell-chsh}), we can see that the state $\tau^{\phi+}$ violates Bell-inequality as $M(\tau^{\phi+})=\frac{1+s^{2}}{2}>1$ implying that the state is entangled $0\leq s\leq 1$. Again from eq.(\ref{bell-chsh}), we see that the state $\tau^{\phi-}$ violates Bell-inequality as $M(\tau^{\phi-})=\sqrt{2}s(1-s)+\frac{1}{2}+s\Big(\frac{3}{2}s-1\Big)>1$ implying that the state is entangled $0\leq s\leq 1$.\\\\
To conclude the analysis of mixture of two qubit state derived from $\vert Star\rangle$ and Bell states, we next consider the states $\tau^{(2)}$ defined in eq.(\ref{class2a}). The density matrices of mixture of $\rho^{star}_{BC}$ with Bell states $\vert \psi^{+}_{BC}\rangle$ and $\vert \psi^{-}_{BC}\rangle$ are given respectively as follows:
\begin{eqnarray}
\label{nonxstate4}
\mathcal{\tau^{\psi+}}=\begin{pmatrix}
2(\alpha-\beta)  & \alpha-\beta & 0 & \alpha-\beta\\
\alpha-\beta & \alpha+\beta  & 2\beta & \alpha-\beta\\
0 & 2\beta & 2\beta & 0\\
\alpha-\beta & \alpha-\beta & 0 & \alpha-\beta\\
\end{pmatrix},\:\:
\mathcal{\tau^{\psi-}}=\begin{pmatrix}
2\beta  & \beta & 0 & \beta\\
\beta & 2\alpha-\beta  & -2(\alpha-\beta) & \beta\\
0 & -2(\alpha-\beta) & 2(\alpha-\beta) & 0\\
\beta & \beta & 0 & \beta\\
\end{pmatrix}.
\end{eqnarray}
Next we proceed for similar analysis of teleportation fidelity of states $\tau^{(2)}$ as done for states $\tau^{(1)}$. With minor moderations in the terms of the density matrices defined in eq.(\ref{nonxstate4}) it can easily be shown that $C(\tau^{\phi^{\pm}})=C(\tau^{\psi^{\pm}})$. Likewise it is observed that $L(\tau^{\phi^{\pm}})=L(\tau^{\psi^{\pm}})$ and $f^{T}(\tau^{\phi^{\pm}})=f^{T}(\tau^{\psi^{\pm}})$ as well.\\\\
We now plot below the teleportation fidelities of the bipartite mixture of states that were defined in sec.\ref{sec:bipartite} such as $\rho^{(1)}(and\:\rho^{(2)})$ from eq.(\ref{class1a}), $\rho^{(3)}(and\:\rho^{(4)})$ from eq.(\ref{class1b}), $\rho^{(5)}(and\:\rho^{(6)})$ from eq.(\ref{class1c}) and $\tau^{(1)}(and \: \tau^{(2)})$ from eq.(\ref{class2a}) by taking a common framework for comparison of teleportation fidelities of the given states, for which we have varied a common state parameter (say $p$) along the horizontal axis from $0$ to $1$. The states of eqs.(\ref{class1a}) and (\ref{class1b}) are of $X$ type and those of eqs.(\ref{class1c}) and (\ref{class2a}) are of non-$X$ types. It is also to be remembered that as $\vert \bar{W}\rangle$ is the spin flipped version of $\vert W\rangle$, and as $\vert W\bar{W}\rangle$ is the state constructed as linear superposition of $\vert W\rangle$ and $\vert \bar{W}\rangle$ and although  $\vert W\bar{W}\rangle$ is of non-$X$ type state while both $\vert W\rangle$ and $\vert \bar{W}\rangle$ are of $X$ type, yet the nature of the teleportation fidelities of these states are similar. In the following figure we plot teleporatation fidelities of $\rho^{(1)}$, $\rho^{(6)}$, $\tau^{\phi+}$ and $\tau^{\phi-}$ to compare how these states behave as quantum teleportation channels.
\begin{figure}[h]
\label{fig2}
\includegraphics[width=16.20cm]{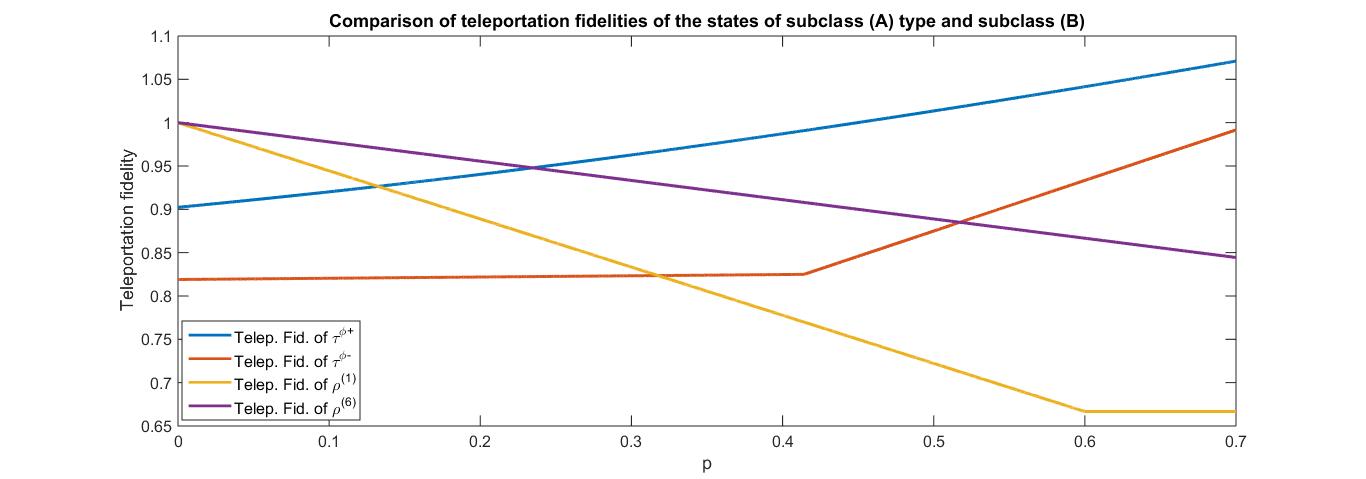}
\caption{The figure represents the teleportation fidelities of the states of subclass (A) and subclass (B) The states $\tau^{\phi+}$, $\tau^{\phi-}$, $\rho^{(1)}$ and $\rho^{(6)}$ have been plotted against common state parameter $p$ ranging from $0$ to $0.45$.}
\label{fig1}
\end{figure}\\
It is being observed from fig.$3$ that the states of subclass (A) and subclass (B) are both useful as quantum teleportation channels. The states $\rho^{(1)}$ of subclass (A) is a non-maximally entangled mixed state of $X$ type whereas the states $\tau^{\phi+}$, $\tau^{\phi-}$ and $\rho^{(6)}$ are non-maximally entangled mixed state of non-$X$ type. The state $\rho^{(6)}$ which has been constructed by taking convex mixture of two qubit state derived from $\vert W\bar{W}\rangle$ and two qubit singlet state performs better as quantum teleportation channels than the state $\tau^{\phi+}$ (which is a convex mixture of two qubit state  derived from $\vert Star\rangle$ and Bell state) for $0\leq p\leq 0.25$ and again for $0.25<p\leq 0.45$. The state $\tau^{\phi+}$ proves to be better candidate as the teleportation channel than $\rho^{(6)}$. However the state $\rho^{(6)}$ is always a better performer than the state $\rho^{(1)}$ (of subclass (A)) as teleportation channel. On the other hand, the state $\tau^{\phi-}$ of eq.(\ref{class2a}) also acts as a better candidate as quantum teleportation channel than the states $\rho^{(1)}$ and $\rho^{(6)}$ when $p>0.5$.\\\\
Finally we also observe that all the non-maximally entangled mixed states defined in this paper outperforms MEMS $\rho^{W}$ and $\rho^{\bar{W}}$ as quantum teleportation channels (note that $f^{T}(\rho^{W})=f^{T}(\rho^{\bar{W}})=\frac{7}{9}$). We also plot in the following the teleportation fidelity of the states $\tau^{\phi+}$, $\tau^{\phi-}$ and $\rho^{(werner)}$ against the common state parameter $\frac{1}{2}\leq p \leq 1$, since in this range the teleportation fidelity of Werner state exceeds classical limit of $\frac{2}{3}$.
\begin{figure}[h]
\label{fig3}
\includegraphics[width=16.20cm]{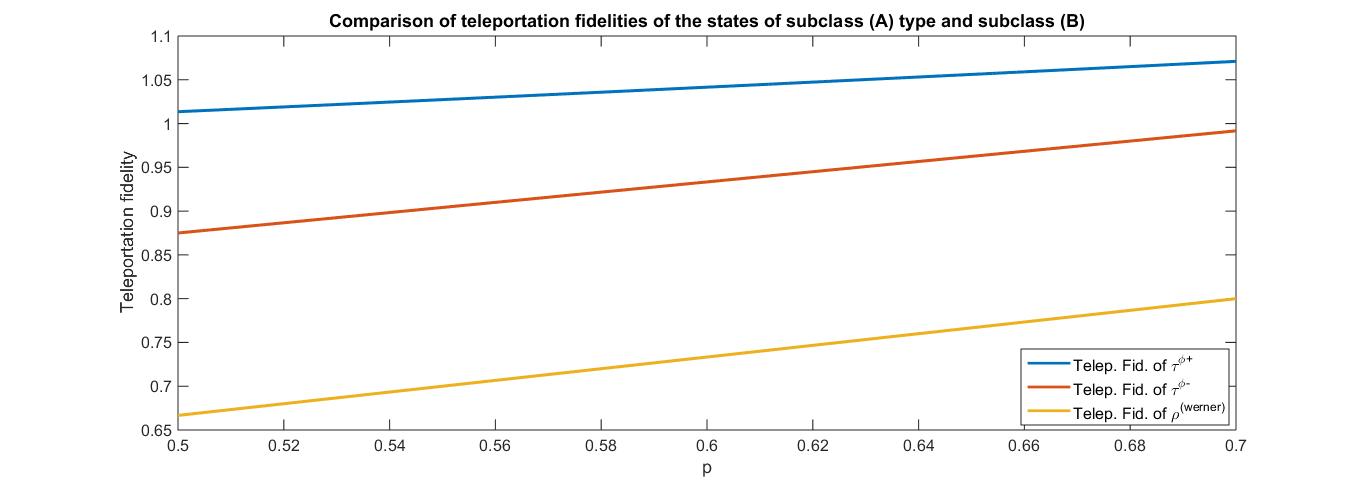}
\caption{The figure shows the plot of teleportation fidelities of the states $\tau^{\phi+}$ , $\tau^{\phi-}$  and $\rho^{(werner)}$, set against the parameter $p$.} 
\label{fig1}
\end{figure}\\
We see from the fig.$3$ that the teleportation fidelities of $\tau^{\phi+}$ and $\tau^{\phi-}$ surpasses the teleportation fidelity of Werner state for $p>0.5$. This implies that the teleportation fidelities of non-maximally entangled mixed states of eq.(\ref{class2a}) outperform that of maximally entangled mixed state such as Werner state. We know that Werner state and $\rho^{W}$ of eq.(\ref{mems1}) are both Ishizaka-Hiroshima class of maximally entangled mixed states. Werner state outperforms $\rho^{W}$ when $p>0.6$ which is also clear from fig.$4$. The class of NMEMS of non-$X$ type defined in eq.(\ref{class2a}) outperforms both of these MEMS of Ishizaka-Hiroshima class.
\section{Conclusion:}
\label{sec:conclusion}
To summarize, in this paper we have studied the efficiency of a few non-maximally entangled mixed states (NMEMS) as resources for quantum teleportation. Horodecki \textit{et.al} showed that there exist inseparable states which are not useful for teleportation within the standard scheme. It was also proposed that the states which violate generalized Bell-CHSH inequality are useful for teleportation\cite{horohorohoro1996}. Later Adhikari \textit{et.al} had shown that not all maximally entangled mixed states (MEMS) are useful for teleportation. Munro class of states is one such example. These states are not useful as quantum teleportation channel when their mixedness exceeds a certain bound while Werner state is another type of maximally entangled mixed states of Ishizaka-Hiroshima type, which are although less entangled for a given degree of mixedness than Munro class, could be more useful as a quantum teleportation channel\cite{adhikari2010}. Moreover Werner states are such states which do not violate Bell-CHSH inequality but are entangled. Adhikari \textit{et.al} also proposed a class of NMEMS states and had shown that they could perform as quantum teleportation channels in spite of satisfying Bell-CHSH inequality. It has been found that, magnitude of entanglement of the states and violation of Bell-CHSH inequality by the states are both not good indicators of their capacity to perform as quantum teleportation channels\cite{adhikari2010}. In this light, we have proposed here a few class of non-maximally entangled mixed states and explored their ability to act as quantum teleportation channels and to know how they behave with respect to Bell-violation.\\\\ For this we have constructed mixtures of class of Bell states with two qubit mixed states derived from three qubit pure states like $\vert W\rangle,\: \vert \bar{W}\rangle$, $\vert W\bar{W}\rangle$ and $\vert Star\rangle$ by removing parties. It is found that some mixtures are of $X$ type and others are not of $X$ type. Our motivation in this work is to explore the characteristics of these two different types of states from the perspective of their usefulness as quantum teleportation channels and how they behave with respect to Bell-CHSH inequality violation. The states $\rho^{(1)}$ (or $\rho^{(2)}$) and $\rho^{(3)}$ (or $\rho^{(4)}$)  are of similar nature as both of these class of states have been obtained as convex combination of two qubit state derived from $\vert W\rangle$ and $\vert \bar{W}\rangle$ respectively and Bell states where $\vert\bar{W}\rangle$ is the spin flipped version of $\vert W\rangle$. Moreover the states $\rho^{(1)}$ (or $\rho^{(2)}$) and $\rho^{(3)}$ (or $\rho^{(4)}$) belong to class of $X$ type states. On the other hand the state $\rho^{(5)}$ (or $\rho^{(6)}$), which has been constructed as mixtures of two-qubit states derived from $3-$ qubit $\vert W\bar{W}\rangle$ and Bell states are found not only to be useful as teleportation channels but also one of such non-$X$ type states viz. $\rho^{C^{2a}}$ of eq.(\ref{nonxstate2}) have higher teleportation fidelity for specific range of state parameter than that of $\rho^{(1)}$ (or $\rho^{(2)}$) and $\rho^{(3)}$ (or $\rho^{(4)}$). We know that tripartite $\vert GHZ\rangle$ and $\vert W\rangle$ are two different SLOCC class of states. We have also constructed mixtures of two qubit state derived from $\vert GHZ\rangle$ and Bell states, which we have designated in the paper by the notation $\rho^{g}$ and when compared with states like $\rho^{(1)}$ (and $\rho^{(2)}$), $\rho^{(3)}$ (and $\rho^{(4)}$) and $\rho^{(5)}$ (and $\rho^{(6)}$), it is observed that the teleportation fidelity of the state $\rho^{g}$ is less than that of the rest for certain range of state parameter. It is to be noted that $\rho^{g}$ is $X$ type states. It is interesting to note down that the teleportation fidelities of these bipartite non-maximally entangled mixed states $\rho^{(1)}$ (and $\rho^{(2)}$), $\rho^{(3)}$ (and $\rho^{(4)}$) (of $X$ type) and $\rho^{(5)}$ (and $\rho^{(6)}$) (of non-$X$ type) exceed the teleportation fidelity of MEMS i.e. $f^{T}(\rho^{W})$ (or $f^{T}(\rho^{\bar{W}})$) of eq.(\ref{mems1}). Bipartite NMEMS obtained from mixtures of two qubit states derived from tripartite $\vert Star\rangle$ and Bell states (denoted as $\tau^{\phi+}$ (and $\tau^{\phi-}$)) are good candidates for quantum teleportation task. First of all, these states are of non-$X$ type. Secondly, unlike the other states which are similar with respect to whichever qubit is removed from the tripartite counterpart, to build states like $\tau^{\phi+}$ (and $\tau^{\phi-}$)), only the peripheral qubits can be removed from the tripartite $\vert Star\rangle$ and not the central qubit (as removal of central qubit makes the states separable). It is found however that the states $\tau^{\phi+}$ (and $\tau^{\phi-}$)) give better performance as quantum teleportation channels than those of $\rho^{(1)}$ (and $\rho^{(2)}$), $\rho^{(3)}$ (and $\rho^{(4)}$) and $\rho^{(5)}$ (and $\rho^{(6)}$) and $\rho^{g}$. The states  $\tau^{\phi+}$ (and $\tau^{\phi-}$)) are NMEMS and we also observe that the teleportation fidelities of such NMEMS exceed the teleportation fidelity of Werner state $\rho^{(werner)}$ (which is a MEMS).\\\\
The study of Bell-CHSH inequality violation by the states is another aspect which have been looked upon in this work. Althoughit is known that a state indicating Bell-violation is suitable as teleportation channel\cite{horohorohoro1996}, yet there are states which are entangled but satisfy Bell-inequality and are also useful for teleportation \cite{adhikari2010}. The states $\rho^{(1)}$ (and $\rho^{(2)}$), $\rho^{(3)}$ (and $\rho^{(4)}$) of eqs.(\ref{class1a}) and (\ref{class1b}) of $X$ type satisfy Bell-inequality and yet they are entangled, shown to be useful as teleportation channels for specific ranges of state parameter. In case of states of non-$X$ type, that we have constructed, behave differently. The states $\rho^{(5)}$ (and $\rho^{(6)}$) of eq.(\ref{class1c}) are examples of states which satisfy Bell-inequality for specified ranges of state parameter, yet they are entangled and useful for teleportation whereas another non-$X$ types of states $\tau^{\phi+}$ (and $\tau^{\phi-}$)) are found to violate Bell-inequality. Consequently these states are entangled can be useful as quantum teleportation channels.\\\\
In future, we can analyze the decoherence effects on these NMEMS states. The experimental realization of these non-maximally entangled mixed states can also be studied.

\section*{Data Availability Statement}
The authors confirm that the data supporting the findings of this study are available within the article.

\end{document}